\documentclass[preprintnumbers,amsmath,11pt,amssymb,floatfix,superscriptaddress,nofootinbib]{article}
\usepackage{relsize,exscale}
\def\mytitle#1{\setcounter{equation}{0}
\setcounter{footnote}{0}
\begin{flushleft}\Large\textbf{#1}\end{flushleft}
\vspace{0.25cm}}
\def\myname#1{\leftline{{\large #1}}\vspace{-0.13cm}}
\def\myplace#1#2{\small\begin{flushleft}\textit{#1}\\
\texttt{#2}\end{flushleft}}

\def\myclassification#1{\small\noindent
Keywords :
       #1\vspace{0.5cm}}
\usepackage{graphicx}
\usepackage[square,numbers]{natbib}
\usepackage{amsmath}
\usepackage{bigints}
\usepackage{cleveref}
\begin{document}
\mytitle{Accretion of Dark Energy Candidates Following Redshift Parametrization Type Equation of State : Horava-Lifshitz Gravity}

\myname{Promila Biswas * \footnote{promilabiswas8@gmail.com}, Sukanya Dutta** \footnote{sduttasukanya@gmail.com}, Ritabrata Bisawas** \footnote{biswas.ritabrata@gmail.com} and Farook Rahaman* \footnote{farookrahaman@gmail.com}}
\myplace{*Department of Mathematics, Jadavpur University, Kolkata-32, India\\ **Department of Mathematics, The University of Burdwan, Burdwan-713104, India} {}
 
\begin{abstract}
In this article, accretion of particular dark energy candidates is studied. These dark energy models possess equation of state dependent on redshift and some free parameters. Central gravitating object for this accretion model is chosen to be the Kehagias Sfetsos black hole sitting in Horava Gravity. Four special redshift parametrization models, viz. Chevallier-Polarski-Linder,  Jassal-Bagala-Padmanabhan, Barboza-Alcaniz and Barboza-Alcaniz-Zhu-Silva are picked to study different accretion properties. Formation of critical point and related radial infalling speed, sonic speed, different arbitrary parameters' values etc are obtained. Mass growth curves are plotted. The rate of growth is followed to depend on the dark matter's behavior as well as the free parameters of dark energy models. Future loss in mass is noted for some of the chosen dark energy models. 
\end{abstract}
\myclassification{Dark energy, Accretion, Cosmology}\\
PACS NO: 95.36.+x , 98.60.Lc  , 98.80
\section{Introduction}

Our universe is found to expand sometimes with a rapidly accelerated speed, sometimes being retarded by gravity and somewhere with a steady accelerating nature. American astronomer Henrietta Swan Leavitt, in 1912, established a relation between the regular luminosity period of a Cepheid star with its absolute magnitude. Using this as a tool, distances of nearby galaxies hosting Cepheid variables were measured. One contemporary scientist of her time, Vesto Slipher noticed redshifts in far away galaxies.

Meanwhile, two publications from Russia(1922) and Belgium(1927) came in front (by Alexander Friedmann and Georges Lemaître respectively) which have shown different upcoming possibilities for the cosmos, especially a tendency to grow a nonstatic future which was contrary to the idea of Einstein's theory.

In 1929, redshifts of several spiral galaxies were measured by Milton Humason. Humason, with his collaborator Edwin Hubble, soon found the distances of Cepheid variables located in these galaxies. Standard luminosity of Cepheid variables allows to measure both the distance and redshift of host galaxies which reside in ``nearby'' region. A linear fit to such a redshift vs. distance data was plotted and the corresponding analysis became popular as Hubble’s law which supported the expansion of cosmos. However, expectations that the speed of the universe will slow down with time were formed.

Another kind of standard candles, type 1a supernovae residing in different farthest galaxies were observed by two teams, captained by astronomers Adam Riess, Saul Perlmutter and Brian Schmidt around 1998 which predicted the late time cosmic acceleration \cite{Perlmutter_1997, Garnavich_1998, Bahcall_1999, Perlmutter_1999, Filippenko_1998}. Justifications towards the question why our universe is expanding faster over time were tried to build via two major pathways. Modification of gravity and inclusion of exotic stress energy.

The first mentioned way led to different modified gravities. General Relativity (GR) was able to perfectly support solar system gravity precision tests , e.g., gravitational redshift, gravitational lensing of distant stars' light, Mercury’s perihelion precision, Shapiro’s time delay effect etc. But in the outer world, some cataclysmic events like black hole-black hole merger \cite{Abbott_2017} have pointed that a quantum correction of GR to justify such strong gravitational effects is required.

A $d$-dimensional manifold $\cal{M}$ invariant under diffeomorphism, equipped with a metric $g_{\mu \nu}$, constructs the base of GR as a field theory. The action of this case looks like 
\begin{equation}\label{action eq}
S_{GR} = -\frac{1}{16 \pi G_N^{(d)}} \int d^d x \sqrt{|g|} \left(\tilde{R} - 2\tilde{\Lambda}\right) + \int d^d x \sqrt{|g|} \cal{L_M}~~~~.
\end{equation}
Here $\tilde{R}$ is scalar curvature of the concerned manifold, $\tilde{\Lambda}$ is the cosmological constant, $g=det(g_{\mu \nu})$ and $\cal{L_M}$ denotes a generic matter Lagrangian. Generalised Newton's constant $G_N^d$ has the energy dimension $[G_N^{(d)}]=2-d$ which reduces to $[G_N^{(4)}]=-2$ in $4D$.

Variating equation \eqref{action eq} with respect to the metric $g_{\mu \nu}$, we obtain the Einstein field equation ($c=1$)
\begin{equation}
\tilde{R}_{\mu \nu} - \frac{1}{2} \tilde{R} g_{\mu \nu} + \tilde{\Lambda} g_{\mu \nu} = 8 \pi G T_{\mu \nu}~~~~,
\end{equation}
here the energy momentum tensor $T_{\mu\nu} = -\frac{2}{\sqrt{|g|}} \frac{\delta~~~~}{\delta g^{\mu \nu}} \left(\sqrt{|g|} \cal{L_M}\right)$. Modification in the Lagrangian led us to different modified gravities.

In the next section (i.e., in $\cref{sec:HLKS}$), we will briefly state about a particular type of black hole solution in Horava Lifshitz(HL) gravity, popularly known as Kehagias Sfetsos solution. In $\cref{sec:RedParDE}$, some special kind of dark energy models will be discussed. In $\cref{sec:DEAccretion}$ dark energy accretion model will be built for KS black hole. In the last section, we will discuss the results in brief.
\section{Horava Lifshitz Gravity and Kehagias Sfetsos Black Hole }\label{sec:HLKS}

The concept of Horava-Lifshitz $(HL)$ gravity incorporates the manifold $\cal{M}$ with a foliation structure, which lets the metric to be written in the form \cite{Blas_2011}
$$ds^2 = N^2 dt^2 - \gamma_{ij}(dx^i+N^i dt)(dx^j+N^j dt)~~~~,$$ where $N$, $N^i$ and $\gamma_{ij}$ are respectively the lapse, shift and spatial leaf metric. 

The said foliation preserves diffeomorphism $t\rightarrow t'(t)$, $x^i \rightarrow x^{i' t}(t,x)$.
For this transformation, $N$, $N^i$ and $\gamma_{i j}$ change as $$N \rightarrow N\frac{dt}{dt'}~~,~~ N^i \rightarrow \left(N^j \frac{\partial x'^{i}}{\partial x^j} - \frac{\partial x'^{i}}{\partial tj} \frac{dt}{dt'}\right)~~~\text{and}~~~\gamma_{ij} \rightarrow \gamma_{kl} \frac{\partial x^k}{\partial x'^{i}}\frac{\partial x^l}{\partial x'^{j}}~~~~~~~~~~~.$$

Extrinsic curvature $K_{i j}$ covariantly transforms under these changes as $$K_{ij} = \frac{1}{2N}(\partial_t \gamma_{i j} - \nabla_i N_j - \nabla_j N_i)~~~~.$$

With this, the modified action for $HL$ gravity turns to be 
\begin{equation}
S_{HL} = \frac{1}{16 \pi G} \int dt~ d^Dx~ N \sqrt{\gamma} [K_{ij}K^{ij} - \lambda K^2 - \cal{V}]~~~~,
\end{equation} 
where $\sqrt{|g|}=N\sqrt{\gamma}$, $D=d-1$ is the dimension of spatial leafs, $\lambda$ is a dimensionless coupling and $\cal{V}$ contains only spatial derivatives. $K_{ij}$ is the only covariant tensor which involves time derivatives. This makes the action second order in time and let to avoid the presence of Orstrogradsky ghosts for all observers. If ${\cal V} = R - 2\Lambda$ and $\lambda=1$, the action noted reproduces the $D+1$ decomposition of the Einstein-Hilbert action with the symmetry group enhanced to full diffeomorphisms.

For a detailed balance conditioned $HL$ gravity, the action takes the explicit form 
$$S = \int dt~d^3x~\sqrt{g}N\left[\frac{2}{\kappa^2}(K_{lm}K^{lm}-\lambda K^2) - \frac{\kappa^2}{2 w^4} C_{lm}C^{lm} + \frac{\kappa^2 \mu \epsilon^{lmt}}{2 w^2} R_{lq} \nabla_m R^q_l \right.$$ 
\begin{equation}
\left. - \frac{\kappa^2 \mu^2}{8} R_{lm} R^{lm} + \frac{\kappa^2 \mu^2}{8(1-3\lambda)} \left(\frac{1-4\lambda}{4} R^2 + \Lambda R -3\Lambda^2\right) + \mu^4 R \right]~~~~,
\end{equation}
where $w$, $\kappa$ and $\lambda$ symbolize coupling constants which are dimensionless. $\Lambda$ is the positive cosmological constant related to the Infrared limit and posessing mass dimension $2$. An antisymmetric tensor $\epsilon^{lmt}$ and a parameter $\mu$ with mass dimension $1$ are introduced. The extrinsic curvature $K_{lm}$ is now written with respect to the metric tensor
\begin{equation}
K_{lm} = \frac{1}{2N} (\dot{g}_lm-\nabla_l N_m - \nabla_m N_l)~~~~.
\end{equation}
Cotton tensor $C^{lm}= \epsilon^{ltq} \nabla_t (R^m_q - \frac{1}{4} \delta^m_q).$ Using the projectibility condition of $HL$ gravity \cite{Thomas_P_Sotiriou_2009} and choosing the Friedmann-Robertson-Walker metric as $N=1$, $g_{lm}=a^2(t) \gamma_{lm}$, $N^l=0$ and 
\begin{equation}
\gamma_{lm} dx^l dx^m = \frac{dr^2}{1-\kappa r^2} + r^2 d \Omega^2_2~~~,
\end{equation}
where the curvature parameter $K$ can opt the values $-1,~ 0$ $\& $ $1$ when it specifies open, flat or closed universe respectively. $a(t)$ is the scale factor.

Choosing dark matter to take the form of a perfect fluid, the related Friedmann equation transforms into 
\begin{equation}
H^2=\frac{{\cal{\kappa}}^2}{6(3\lambda-1)}\left[\rho_{tot} - \frac{6 k \mu^4}{a^2} - \frac{3 k {\cal{\kappa}}^2 \mu^2}{8(3\lambda-1)a^4}\right]~~~~.
\end{equation}
Here $H=\frac{\dot{a}}{a}$ is the Hubble parameter and $\rho_{tot}$ is usual energy density $=\rho_m + \rho_{\phi}$. A Schwarzschild like solution was obtained in HL gravity by Kehagias and Sfetsos \cite{Culetu_2015}. Considering $\lambda =1$ the fluid equation turns 
\begin{equation}
H^2=\frac{{\cal{\kappa}}^2}{12}\left[\rho_{tot} - \frac{6 k \mu^4}{a^2} - \frac{3 k {\cal{\kappa}}^2 \mu^2}{16 a^4}\right]~~~,
\end{equation}
when a flat space is chosen, i.e., the curvature parameter $k=0$, higher order derivative terms will give no contribution in action. However, for a nonzero $k$ significant changes are expected with small value of the scale factor $a(t)$.

The KS solution is given by \cite{Abbas_2013}
\begin{equation}
ds^2 = A(r) dt^2 - \frac{1}{A(r)} dr^2 - r^2 (d\theta^2 + sin^2 \theta ~d\phi^2)~~~~,
\end{equation}
where $A(r)$ is given as 
\begin{equation}
A(r) = 1+ w r^2 - w r^2 \sqrt{1+ \frac{4 M}{w r^3}}~~~~,
\end{equation}
where $w = \frac{16 \mu^2}{{\cal{\kappa}}^2}.$

 In section $\cref{sec:RedParDE}$, we will now recapitulate regarding some dark energy candidates for which equation of states are functions of redshift.
\section{Redshift Parametrization Models As Dark Energy Candidates}\label{sec:RedParDE}
We will shift our eye towards the modification of stress energy side of field equation. This is commonly known as the involvement of dark energy (DE). Several candidates are proposed. `Quintessence' is the first kind of candidate where equation of state looks like \cite{Peebles_1987}
\begin{equation}
w_Q = \frac{\frac{1}{2}\dot{\phi^2}-V(\phi)}{\frac{1}{2}\dot{\phi^2}+V(\phi)}~~~,
\end{equation}
where $\phi$ and $V$ are respectively related scalar field and potential. 

`Phantom' energy is another kind of model \cite{Caldwell_2003} which generates quick acceleration leading to future Cosmic singularity named ``Big Rip''.

Combining quintessence and phantom, another hypothetical scenario regarding quintom DE model is proposed. For low redshift, EoS looks like,
\begin{equation}
w_{DE}(z) = w_0 +w^{'}(z)~~~,
\end{equation}
which, for high redshift, turns 
\begin{equation}
w_{DE}(z)= w_0 +w_1 (1-a)= w_0 + w_1 \frac{z}{1+z}~~.
\end{equation}

Free parameters of this model are determined as $w_0 =1.06 \pm 0.14$ and $w_1=0.36 \pm 0.62$ \cite{Serra_2011}. This model can not account for why cosmic single fluid model or a scalar field model fails to realise available Quintom model in FLRW universe. To justify, either a non-minimal coupling is to be introduced or extra degrees of freedom will be required \cite{Cai_2010}. 

Next generation DE models are proposed as 
$$w(z) \approx w_0 +  \left. \left(\frac{d w}{dz} \right) \right|_{z=0} \times z ~~~~.$$ 

Amount of acceleration will be high if $\left. \left(\frac{d w}{dz} \right) \right|_{z=0}$ is negative. Assuming a flat universe and applying data from SDSS, dimensionless density for dark matter, $\Omega_{DM} = 0.3 \pm 0.1$, EoS is found to be $ w(z) \sim -1$. Linear Redshift Parameterization (LRP) \cite{Cooray_1999} has the equation of state $$w^{LRP}(a) = w_0^{LRP} + w_1^{LRP} \left( \frac{1-a}{a}\right)~~~~.$$ 

In the article \cite{Chevallier:2000qy}, Chevallier-Polarski-Linder (CPL) Redshift parameterization model is proposed which is able to reproduce the cosmic picture for $z \leq 2$. At $z=0$, $\Lambda$CDM is almost perfectly mimicked. A special care has been taken where Cosmic Microwave Background (CMB) peak is formed. EoS is given as 
\begin{equation}\label{EoS of CPL}
w^{CPL}(z) = w^{CPL}_0 + w^{CPL}_1 \frac{z}{1+z}~~~.
\end{equation}
Jassal-Bagala-Padmanabhan (JBP) Redshift Parameterization \cite{Jassal_2004} proposed a model with the EoS 
\begin{equation}\label{EoS of JBP}
w^{JBP}(z) = w^{JBP}_0 + \frac{w^{JBP}_1~z}{(1+z)^2}~~~~.
\end{equation}

This model is shown to restrict huge variation of energy density at low redshift when SNeIa observations with the constraints from WMAP observation are applied. 

In the article \cite{Barboza_2008}, Barboza-Alcaniz (BA) redshift parameterization EoS is proposed as, 
\begin{equation}\label{EoS of BA}
w(z) = w_0^{BA} + w_1^{BA} \frac{z(1+z)}{1+z^2}~~~~~~~~~,
\end{equation}
while constraining free parameters, we find for $68.3\%$ confidence, $w_0= -1.14_{-0.24}^{+0.31}$, $w_1 = 0.84_{-1.59}^{0.65}$ and $\Omega_m^0=0.27 \pm 0.003$. Cosmos depicted in this model is found to transit from deceleration to acceleration at $z_t \simeq   0.58$.  

Barboza-Alcaniz-Zhu-Silva (BAZS) \cite{Barboza_2009} have considered another parameterization
$$\omega^{BAZS}(a) = \omega_0 - \omega_{\beta} \frac{a^{\beta}-1}{\beta}~~~~,$$ i.e.,
\begin{equation}\label{EoS of BAZS}
\omega^{BAZS}(z)= \omega_0^{BAZS} - \omega_{\beta} \frac{(1+z)^{-\beta}-1}{\beta}~~~~~.
\end{equation}
We have constrained three free parameters, we have found three kinds of confidence contour graphs and the best fit values of $w_0^{BAZS}$, $w_{\beta}$ and $\beta$.

We will solve different fluid's densities using the cosmological equation of continuity
\begin{equation}\label{conservation eqn for fluid}
\dot{\rho}=-3H\left(\rho+p\right)~~~~~.
\end{equation}

Energy density of all the said DE models can be determined. In the next section, we will study the accretion of different DE candidates upon Kehagias Sfetsos black holes.
\section{Dark Energy Accretion Towards Kehagias Sfetsos Black Holes}\label{sec:DEAccretion}
Assuming the fluid to follow perfect fluid nature and recalling the  energy momentum tensor of DE 
\begin{equation}
    T_{\mu \nu}=(\rho +p)u_\mu u_\nu +p g_{\mu \nu}~~~~,
\end{equation}
$\rho$ and $p$ respectively are the energy density and the pressure of DE. For the present work, we will choose $\rho=\rho_{\phi}^{i}$, $p=p^i_{\phi}$, where `$i$' will tell exactly which DE model is chosen. $u^\mu\equiv \frac{dx^\mu}{d\tau}$ is the four speed for the $\mu^{th}$ coordinate, $x^\mu$. $\tau$ is proper time. 4-velocity follows the norm property, $u_\mu u^\nu=1$. We also suppose that the black hole's spherical symmetry is not influenced by the accreted DE.

Without any loss of generality, we assume $u^1=u<0$ (as this is radially infalling) and $u^2=u^3=0$ due to symmetry and  velocity components are derived as 
$$    u_1=\frac{u}{{\cal{A}}(r)},~~u^0=\frac{\sqrt{{\cal{A}}(r)+u^2}}{{\cal{A}}(r)}~~\text{and}~~ u_0=-\sqrt{h(r)+u^2}~~~~~.$$
To construct the continuity equation, we consider the components of $T^{\nu}_{\mu}$ as
\begin{equation}
 T^0_0=(\rho + p)u^0 u_0 + p =-\rho - \frac{u^2}{h(r)} (\rho + p)~,~T^1_1 = (\rho + p) u^1 u_1 + p =(\rho + p)\frac{u^2}{h(r)} + p~,~T^2_2=T^3_3=p~~~\text{and}~~~T_0^1 = -u(\rho + p) \sqrt{h(r) +u^2} ~~.  
\end{equation}
Along with the 4-current, conservation of mass flux $\frac{\partial J^\mu}{\partial x^\mu} = 0$ can be written as
\begin{align}
    J^\mu = \rho \begin{pmatrix} c \\ \dot{x}^i \end{pmatrix} = \begin{pmatrix} c \\ u \\ 0 \\0 \end{pmatrix} 
    \implies \rho~ u~ r^2 = \lambda_0 \text{ (a~ constant)}
\end{align}
and opting the $0^{th}$ component of $T^{\nu}_{\mu ; \nu} = 0$, we obtain 
\begin{equation}\label{components of T mu nu}
    0 = T^{\nu}_{0;\nu} = T^0_{0 ; 0} + T^1_{0 ; 1} + T^2_{0 ; 2} + T^3_{0 ; 3}
\implies u r^2 (\rho + p) \sqrt{{\cal{A}}(r)+u^2}M^{-2}=\lambda_1~~~~,
\end{equation}
where $\lambda_1$ is another arbitrary constant of integration.

Next we will consider the energy momentum conservation $u^\mu T^\nu_{\mu ; \nu} =0$, i.e., the projection of energy momentum conservation law on one fluid's 4-velocity and we obtain \cite{Debnath_2015}
\begin{equation}\label{fluid four velocity}
    ur^2 M^{-2} exp \left\{ \int^\rho_{\rho_\infty} \frac{d {\rho}'}{{\rho}' + p({\rho}')}\right\} = - \lambda_2~,
\end{equation}
$\lambda_2 > 0$ symbolizes another integrating constant related the energy flux \cite{Debnath_2015, Babichev_2004, Babichev_2005, Babichev_2013}. $\rho$ and $\rho_{\infty}$ represent the values of the density at finite distance and at infinity respectively.

We choose a dimensionless radial distance parameter $x=\frac{r}{M}$, $M$ is the mass of central gravitating object. Hence equation \eqref{fluid four velocity} reduces to 
\begin{equation}\label{lambda 2}
   ux^2 exp \left\{ \int^\rho_{\rho_\infty} \frac{d {\rho}'}{{\rho}' + p({\rho}')}\right\} = - \lambda_2~.
\end{equation}

Now we choose a formal arbitrary function $n$ such that 
\begin{equation}
    \frac{d \rho}{\rho +p} = \frac{d n}{n}~~~~~.
\end{equation}

Here the medium is chosen to possess only individual conserved particles \cite{Babichev_2005} and the corresponding number density is taken as $n$.

Integrating equation \eqref{lambda 2} we obtain
\begin{equation}\label{n by n infinity}
    \frac{n}{n_\infty} \equiv exp \left\{ \int^\rho_{\rho_\infty} \frac{d {\rho}'}{{\rho}' + p({\rho}')}\right\}~~,
\end{equation}
where $n_\infty$ is called the concentration of DE \cite{Babichev_2005}.

From equation \eqref{components of T mu nu}, we obtain 
\begin{equation}\label{lambda 1}
    u x^2 (\rho +p) (\sqrt{u^2 + \cal{A}}) = \lambda_1
\end{equation}

This converts equation \eqref{lambda 1} into 
\begin{equation}\label{lambda equations}
    \frac{\rho + p}{n} \sqrt{u^2 + \cal{A}} = \lambda_3 = -\frac{\lambda_1}{\lambda_2} = \frac{\rho_\infty + p(\rho_\infty)}{n_\infty}~~~.
\end{equation}

To determine $\lambda_2$, which has the dimension of energy flux, we evaluate different fluid parameters at the critical point. Let us consider a parameter having the same dimension as the velocity given by
\begin{equation}
    V^2 = \frac{n}{\rho + p} \frac{d (\rho + p)}{dn} -1~~~~.
\end{equation}

We can conclude that $V^2$ is equivalent to the effective sound speed squared of the concerned accreting medium. We symbolize and define it as 
\begin{equation}
    V^2 = c^2_s (\rho) = \frac{\partial p}{\partial \rho}~~~~.
\end{equation}

Now differentiating equations \eqref{lambda 2} and \eqref{lambda equations} and using \eqref{n by n infinity}, we obtain 
\begin{equation}
    \frac{du}{u} \left[V^2 -\frac{u^2}{u^2 + \cal{A}}\right] + \frac{dx}{x} \left[2V^2 - \frac{x {\cal{A}}'}{2(u+\cal{A})}\right]=0~~~~.
\end{equation}

Clearly the radial inward speed gradient $\frac{du}{dx}$ is now expressed as a numerator to denominator ratio. It is followed, in the interval $0\leq u \leq $ speed of light, the denominator turns zero at some critical distance $x=x_c$.

for a physical flow, the numerator should vanish at the same critical point $x_c$ as well. Hence at $x=x_c$, we obtain two relations 
\begin{equation}\label{uc}
    u_c^2 = \frac{x_c {\cal{A}}'(x_c)}{4}
\end{equation}
and 
\begin{equation}\label{Vc}
    V_c^2 = \frac{x_c {\cal{A}}'(x_c)}{x_c{\cal{A}}'(x_c)+4 {\cal{A}}(x_c)}~~~~.
\end{equation}
where $'\equiv \frac{d}{dx}$ denotes the differentiations with respect to $x$.

Now, using equations \eqref{uc} and \eqref{Vc}, from equation \eqref{lambda 1}, we derive,
\begin{equation}
    \frac{\rho_c +p(\rho_c)}{\rho_\infty + p(\rho_\infty)} = 2\left[ x_c {\cal{A}}'(x_c)+4 {\cal{A}}(x_c) \right]^{-\frac{1}{2}} exp \left\{ \int^{\rho_c}_{\rho_\infty} \frac{d {\rho}'}{{\rho}' + p({\rho}')}\right\}~~~~.
\end{equation}

 Wherever the critical point $x_c$ takes place, we can calculate the density $\rho_c = \rho(x_c)$ there. Again if a particular value of $\rho_c$ is given, we can obtain corresponding $x_c$ and $u_c$.

 Hence it will be very straight forward to calculate 
 \begin{equation}
     \lambda_2 = \frac{x_c^3{\cal{A}}'(x_c)}{2} exp \left\{ \int^{\rho_c}_{\rho_\infty} \frac{d {\rho}'}{{\rho}' + p({\rho}')}\right\}~~~,
 \end{equation}

 So, for different redshift parametrizations, we can obtain the constants $\lambda_2.$

 The rate of change of mass of the black hole is calculated as 
 \begin{equation}
     \dot{M}=-4 \pi \lambda_1 M^2\\
     = 4 \pi \lambda_2 M^2 [\rho_\infty + p (\rho_\infty)]~~~~.
 \end{equation}

 Different references \cite{Abbas_2013, Babichev_2008, Jamil_2010, Jamil_2010_GRG, Dutta_2019} suggests that the rate of change of mass can be calculated at every pair of $\left(\rho, ~p(\rho)\right)$, which does not
satisfy the dominant energy condition, whenever we apply the equation of state $p=w\rho$.

Hence 
\begin{equation}
     \dot{M} = 4 \pi \lambda_2 M^2 [\rho + p (\rho)]~~~~.
\end{equation}
 Sign of $\dot{M}$ is solely determined by that of $(\rho+p)$. For quintessence $-1 < \frac{p}{\rho} < -\frac{1}{3}$ and hence $\dot{M}>0$. But once phantom barrier $(\frac{p}{\rho}= -1)$ is crossed, $p+\rho< 0$ and hence $\dot{M}<0$.

 $$
     \frac{dM}{d\rho} = -\frac{4 \pi \lambda_2 M^2}{3 H} \implies \int^{M_0}_M \frac{dM}{M^2} = -\frac{4\pi \lambda_2}{3}\int^{\rho_0}_{\rho} \frac{d\rho}{H}\implies \frac{1}{M}-\frac{1}{M_0} = -\frac{4\pi \lambda_2}{3}\int^{\rho_0}_{\rho} \frac{d\rho}{H}$$
     
     $$\implies \frac{1}{M} = \frac{1}{M_0} + \frac{4 \pi \lambda_2}{3} \int^{\rho}_{\rho_0}\frac{d\rho}{H}$$
     
      \begin{equation}\implies M = \frac{M_0}{1 + \frac{4 \pi \lambda_2 M_0}{3}\int^{\rho}_{\rho_0}\frac{d\rho}{H}}~~~~~,\end{equation}

      here $M_0~(M(z=0))$ is the present day mass of the black hole. Present time energy density $\rho(z=0)=\rho_0$ is constituted of two parts : present time matter density $\rho_m$ and same for dark energy, i.e., $\rho_{DE_0}$.

      Hence we obtain the equation for 4-D Kehagias-Sfetsos(KS) black hole in Horava-Lifshitz (HL) gravity as 
      \begin{equation}\label{KS BH in HL gravity}
          M = \frac{M_0}{1 + \frac{4 \pi \lambda_2 M_0}{3}\mathop{\mathlarger{\mathlarger{\int}}}^{\rho}_{\rho_0}\frac{d\rho}{\sqrt{\frac{\kappa^2}{12}\left[\rho -\frac{6 k \mu^4}{a^2}-\frac{3 k \kappa^2 \mu^2}{16 a^4} \right]}}}~~~~~~~.
      \end{equation}

 We consider two main contributories to the density and pressure, viz., DE density $(\rho_{DE})$ and matter density $(\rho_m)$.

 Hence,
 $$\rho_{tot} = \rho_{DE}+\rho_m$$  and  $$p_{tot}=p_{DE}+p_m~~~~.$$

 Matter density is solved from \eqref{conservation eqn for fluid} as \cite{Debnath_2015_Astrophys}
 \begin{equation}
     \rho_m = \rho_{m_0} (1+z)^{3(1+w_m)}~~~~~,
     \end{equation}
     where $w_m = \frac{p_m}{\rho_m}$ is the EoS for matter contribution and $\rho_{m_0}~(=\rho_m(z=0))$ is the present time value of matter density. In terms of dimensionless density parameter, $\Omega_{m_0} = \frac{\rho_{m_0}}{3 H_0^2}$, $\rho_m$ can be written as 
     \begin{equation}
         \rho_m = 3 H_0^2 \Omega_{m_0} (1+z)^{3(1+w_m)}~~~~~~.
     \end{equation}

With these details, we are going to study the accretion properties of four different DE models in the next four subsections.
\subsection{Accretion of Dark Energy Following Chevallier-Polarski-Linder Parametrization onto Kehagias-Sfetsos Black Hole}
Solving the equation of continuity \eqref{conservation eqn for fluid} for the CPL EoS \eqref{EoS of CPL}, we obtain the energy density of CPL type DE as 
\begin{equation}\label{rho phi0 of CPL} 
\rho^{CPL}_{\phi} (z) = \rho^{CPL}_{\phi 0}~ exp\bigg\{-3w_1^{CPL}~\frac{z}{1+z}\bigg\}(1+z)^{3\left(1+w^{CPL}_0+w^{CPL}_1\right)}~~~~~~~~,
\end{equation}
where $\rho_{\phi 0}^{CPL}$ is the DE energy density at $z=0$. Hence the total energy density of universe filled up by CPL type of fluid is evolved as
\begin{equation}\label{rho tot of CPL}
\rho^{CPL}_{tot}(z) = 3H_0^2 \left[\Omega_m (1+z)^3 + \Omega^{CPL}_{\phi 0}~ exp\bigg\{-3w_1^{CPL}~\frac{z}{1+z}\bigg\}(1+z)^{3\left(1+w^{CPL}_0+w^{CPL}_1\right)}\right]~~~~.
\end{equation}
So, utilizing the equation \eqref{rho tot of CPL} in the equation \eqref{KS BH in HL gravity}, where $\Omega_{\phi 0}^{CPL}= \frac{\rho_{\phi 0}^{CPL}}{3H_0^2}$ is the dimensionless fractional density for CPL, we get for CPL parametrization,

\begin{equation}
M^{CPL}=\frac{M_0^{CPL}}{1+\frac{4 \pi \lambda_2^{CPL} M_0}{3} \mathop{\mathlarger{\mathlarger{\int}}}_{\rho_{\phi_0}^{CPL}}^{\rho_{\phi}^{CPL}}\frac{d\left(3H_0^2 \left[\Omega_m (1+z)^3 + \Omega^{CPL}_{\phi 0}~ exp\bigg\{-3w_1^{CPL}~\frac{z}{1+z}\bigg\}(1+z)^{3(1+w^{CPL}_0+w^{CPL}_1)}\right]\right)}{\sqrt{\frac{\kappa^2}{12}\left[\left(3H_0^2 \left[\Omega_m (1+z)^3 + \Omega^{CPL}_{\phi 0}~ exp\bigg\{-3w_1^{CPL}~\frac{z}{1+z}\bigg\}(1+z)^{3(1+w^{CPL}_0+w^{CPL}_1)}\right]\right)-6k\mu^4 (1+z)^2-\frac{3k\kappa^2 \mu^2 (1+z)^4}{16}\right]}}}~~~~.
\end{equation} 
The constant $\lambda_2^{CPL}$ is calculated using the equation (\ref{EoS of CPL}) to derive the expression
$$\frac{d\rho'}{\rho'+p(\rho')}=\frac{d\rho'}{\left\{1+w_0+w_1-w_1a(\rho')\right\}\rho'}~~~~.$$
We solve the scale factor $a(\rho')$ numerically as the function of density over a large range. For a Bondi type accretion, we have followed that $x_c$ is likely to occur somewhere inbetween 4.5 to 10 Schwarzschild radius. \cite{Biswas_2011}, \cite{Biswas_2019} Performing numerical integration from a large distant point to such a limited critical value, we have evaluated $\lambda_2^{CPL}$. We have plotted growing mass to zero redshift mass ratio in figure 1a to 1c for different sets of best fit values of the free parameters.
\begin{figure}[h!]
\begin{center}
$$~~~~~~~~~~~~~~~~~Fig-1a~~~~~~~~~~~~~~~~~~~~~~~~~~~~~~~~~~~~~~~
~~~~~~~~~~~~~~~~~~Fig-1b~~~~~~~~~~~~~~~~~~~~~~~~~~~~~~~~~~~~~~~~
~~~~~~~~~~~~~~Fig-1c~~~~~~~~~~~~~~~~~~~~~~~~~~~~~~~~~~~~~$$\\
\includegraphics[height=1.5in, width=2.1in]{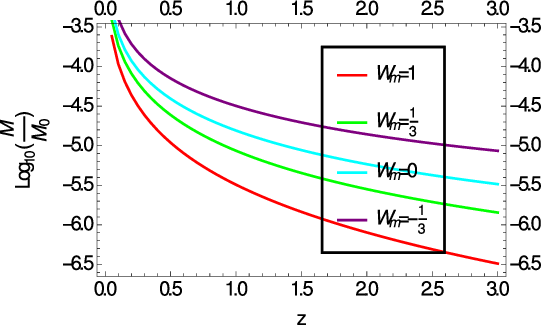}~~~
\includegraphics[height=1.5in, width=2.1in]{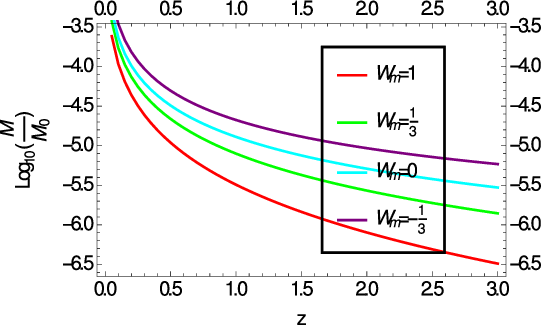}~~~~~
\includegraphics[height=1.5in, width=2.1in]{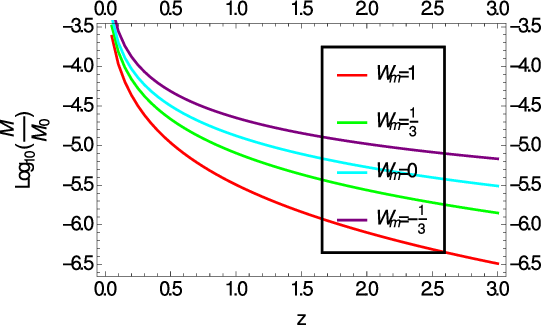}~~~~~\\
\caption{Variations in BH mass $M^{CPL}$ against redshift $z$ due to the accretion of CPL parameterization for different values of $w_m$. We have taken three sets of $w_0=-1$, $w_1=0.81$ in figure 1a, $w_0=-1,~w_1=0.77$ in figure 1b and $ w_0=-0.92,~w_1=0.36$ in figure  1c. The values of $w_0$ and $w_1$ are taken from the article \cite{Promila_thesis_2021}.}
\end{center}
\end{figure}

In figure 1a, where the DE free parameters are chosen as $w_0=-1$ and $w_1=0.81$ \cite{Promila_thesis_2021}, matter equation of state is varied from stiff fluid ($\frac{p}{\rho}=1$) to quintessence barrier ($\frac{p}{\rho}=-\frac{1}{3}$). As $z$ decreases, mass is followed to increase first slowly and then rapidly. Mass at a particular redshift is higher if $w_m$ is low. For a better insight, we have taken logarithm. Shifting to figure 1b and 1c, we observe the same pattern is kept. But near the $z=0$ point, this ratio is higher as the values of $w_1$ is reduced.
\subsection{ Accretion onto Kehagias-Sfetsos Black hole with Dark Energy Following Jassal-Bagala-Padmanabhan (JBP) Redshift Parameterization }
For this candidate, the fractional density contribution turns (using equation \eqref{EoS of JBP} in the equation \eqref{conservation eqn for fluid}).
\begin{equation}\label{rho phi0 of JBP}
\rho^{JBP}_{\phi}(z) = \rho^{JBP}_{\phi 0}~ exp\bigg\{\frac{3w^{JBP}_1}{2(1+z)}\bigg\}(1+z)^{3(1+w^{JBP}_0)}~~~~~~~~,
\end{equation}
where $\rho^{JBP}_{\phi 0} = \rho^{JBP}_{\phi}$ (at $z=0$), comes as the constant of integration.
\begin{equation}\label{rho tot for JBP}
\rho^{JBP}_{tot}(z) = 3H_0^2 \left[\Omega_{m0} (1+z)^{3(1+w_m)} + \Omega^{JBP}_{\phi 0} exp\bigg\{ \frac{3w^{JBP}_1}{2(1+z)}\bigg\}(1+z)^{3(1+w^{JBP}_0)} \right]~~~~~~~,
\end{equation}  
in terms of dimensionless density parameter $\Omega^{JBP}_{\phi 0}=\frac{\rho^{JBP}_{\phi 0}}{3 H_0^2}$.
So, puttting equation \eqref{rho tot for JBP} in the equation \eqref{KS BH in HL gravity}, we will get for JBP parameterization,
\begin{equation}
M^{JBP}=\frac{M_0^{JBP}}{1+\frac{4 \pi \lambda_2^{JBP} M_0}{3} \mathop{\mathlarger{\mathlarger{\int}}}_{\rho_{\phi_0}^{JBP}}^{\rho_{\phi}^{JBP}}\frac{d\left(3H_0^2 \left[\Omega_{m0} (1+z)^3 + \Omega^{JBP}_{\phi 0} exp\bigg\{ \frac{3w^{JBP}_1}{2(1+z)}\bigg\}(1+z)^{3(1+w^{JBP}_0)} \right] \right)}{\sqrt{\frac{\kappa^2}{12}\left[\left(3H_0^2 \left[\Omega_{m0} (1+z)^3 + \Omega^{JBP}_{\phi 0} exp\bigg\{ \frac{3w^{JBP}_1}{2(1+z)}\bigg\}(1+z)^{3(1+w^{JBP}_0)} \right] \right)-6k\mu^4 (1+z)^2-\frac{3k\kappa^2 \mu^2 (1+z)^4}{16}\right]}}}~,
\end{equation}
$M^{JBP}$ is the BH mass in JBP surroundings at an arbitrary redshift $z$ ($M^{JBP}$) and $M_0~(M(z=0))^{JBP}$ is the present mass of the black hole.

We can calculate the constant $\lambda_2^{JBP}$  using
$$\frac{d\rho'}{\rho'+p(\rho')}=\frac{d\rho'}{\left\{1+w_0+w_1a(\rho')\left(1-a(\rho')\right)\right\}\rho'}~~~~.$$
Though the inverse function $a(\rho)$ can not be explicitly determined. We have calculated $\lambda^{JBP}_2$ by the help of numerical integration. $log_{10}\left(\frac{M^{JBP}}{M_0^{JBP}}\right)$ is plotted in figure 2a-c. 
\begin{figure}[h!]
\begin{center}
$$~~~~~~~~~~~~~~~~~Fig-2a~~~~~~~~~~~~~~~~~~~~~~~~~~~~~~~~~
~~~~~~~~~~~~~~~~~~Fig-2b~~~~~~~~~~~~~~~~~~~~~~~~~~~~~~~~~~~~~
~~~~~~~~~~~~~~~~Fig-2c~~~~~~~~~~~~~~~~~~~~~~~~~~~~~~~~~~~~~$$\\
\includegraphics[height=1.5in, width=2.1in]{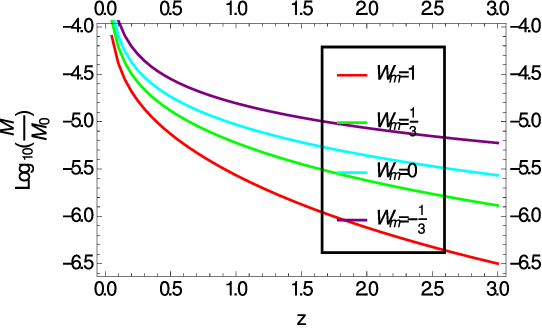}~~~
\includegraphics[height=1.5in, width=2.1in]{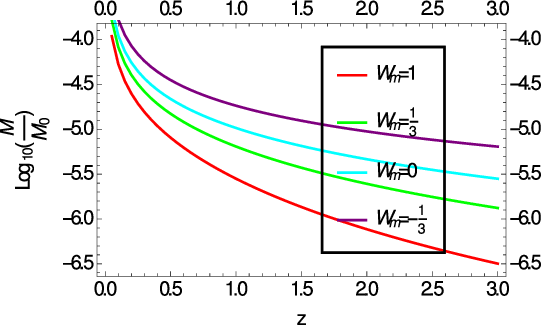}~~~~~
\includegraphics[height=1.5in, width=2.1in]{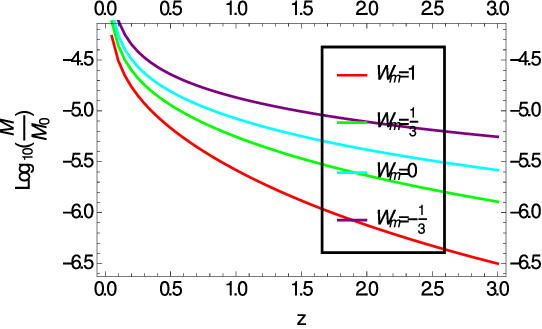}~~~~~\\
\caption{Variations in BH mass $M^{JBP}$ against redshift $z$ due to the accretion of JBP parameterization for different values of $w_m$. We have taken three sets of $w_0$ and $w_1$ values to draw this graph. In figure 1a, $w_0= {-0.75}^{+0.04}_{0.04}$ and $w_1 = {-0.06}^{+0.03}_{-0.03}$. In figure 1b, $w_0={-0.67}^{+0.10}_{-0.10}$ and $w_1 = {-0.19}^{+0.13}_{-0.12}$. In figure 1c, $(w_0, w_1)=({-0.74}^{+0.14}_{-0.16},~{-0.14}^{+0.18}_{-0.17} )$. The values of $w_0$ and $w_1$ are taken from the article \cite{Promila_thesis_2021}.}
\end{center}
\end{figure}
For JBP, near $z=0$, accretion rates are higher than the CPL case. Here if $w_1$ is less, mass at the same redshift is less. Hence accretion is powerful if $w_1$ is high. As CPL, in this case also, when matter part is following lesser value of equation of state accretion is powerful. Even we observe the accretion for $w_m=-\frac{1}{3}$ and $w_m=1$ differ by $\sim 10^2$ of order around $z\sim 3$.
\subsection{Accretion upon Kehagias-Sfetsos Black hole as Accreting Fluid Barboza-Alcaniz (BA) Redshift Parameterization}
Incorporating the equation of state \eqref{EoS of BA} in the cosmological equation of continuity, the fractional distribution of BA parameterization model towards the density turns
\begin{equation}\label{rho phi0 of BA}
\rho^{BA}_{\phi}(z) = \rho^{BA}_{\phi 0} \bigg\{\frac{1+z^2}{(1+z)^2}\bigg\}^{\frac{3w^{BA}_1}{2}} \times (1+z)^{3(w^{BA}_0 + w^{BA}_1)}~~~~~~~~.
\end{equation}
where $\rho_{\phi 0}^{BA}$ is obtained at $z=0$. Hence the density of the concerned model turns $\Omega^{BA}_{\phi 0}=\frac{\rho^{BA}_{\phi 0}}{3 H_0^2}$.
\begin{equation}\label{rho tot for BA}
\rho^{BA}_{tot}(z) = 3H_0^2 \left[\Omega_{m0} (1+z)^{3(1+w_m)} + \Omega^{BA}_{\phi 0} \bigg\{\frac{1+z^2}{(1+z)^2}\bigg\}^{\frac{3w^{BA}_1}{2}} \times (1+z)^{3(w^{BA}_0 + w^{BA}_1)}\right]~~~~~~~~,
\end{equation} 
So, incorporating equation \eqref{rho tot for BA} in the equation \eqref{KS BH in HL gravity}, we will get for BA parameterization,
\begin{equation}
M^{BA}=\frac{M_0^{BA}}{1+\frac{4 \pi \lambda^{BA}_1 M_0}{3} \mathop{\mathlarger{\mathlarger{\int}}}_{\rho_{\phi_0}^{BA}}^{\rho_{\phi}^{BA}}\frac{d\left(3H_0^2 \left[\Omega_{m0} (1+z)^3 + \Omega^{BA}_{\phi 0} \bigg\{\frac{1+z^2}{(1+z)^2}\bigg\}^{\frac{3w^{BA}_1}{2}} \times (1+z)^{3(w^{BA}_0 + w^{BA}_1)} \right] \right)}{\sqrt{\frac{\kappa^2}{12}\left[\left(3H_0^2 \left[\Omega_{m0} (1+z)^3 + \Omega^{BA}_{\phi 0} \bigg\{\frac{1+z^2}{(1+z)^2}\bigg\}^{\frac{3w^{BA}_1}{2}} \times (1+z)^{3(w^{BA}_0 + w^{BA}_1)} \right] \right)-6k\mu^4 (1+z)^2-\frac{3k\kappa^2 \mu^2 (1+z)^4}{16}\right]}}}~~,
\end{equation} 
$M^{BA}$ is the black hole mass embedded in BA at an arbitrary redshift ($M^{BA}$) and $M_0^{BA}~(M^{BA}(z=0))$ is the present mass of the black hole in BA.
\begin{figure}[h!]
\begin{center}
$$~~~~~~~~~~~~~~~~~~~~~~~~~~~~~Fig-3a~~~~~~~~~~~~~~~~~~~~~~~~~~~~~~~~~~~~~~~~~~~~~~~~
~~~~~~~~~Fig-3b~~~~~~~~~~~~~~~~~~~~~~~~~~~~~~~
~~~~~~~~~~~~~~~~~~~Fig-3c~~~~~~~~~~~~~~~~~~~~~~~~~~~~~~$$\\
\includegraphics[height=1.5in, width=2.1in]{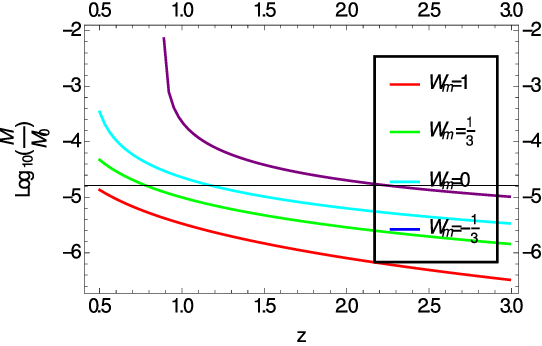}~~~
\includegraphics[height=1.5in, width=2.1in]{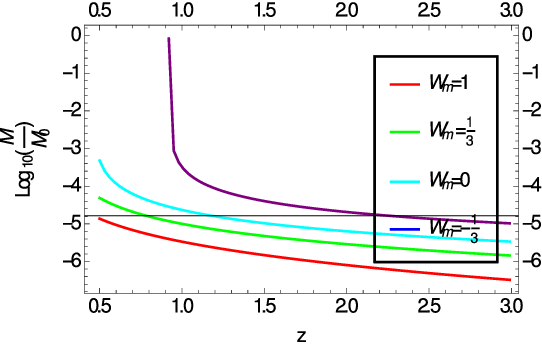}~~~~~
\includegraphics[height=1.5in, width=2.1in]{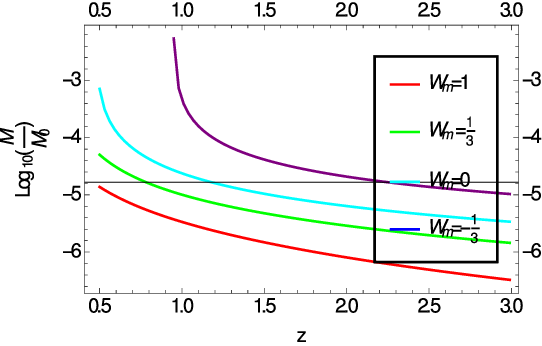}~~~~~\\
\caption{Variations in BH mass $M^{BA}$ against redshift $z$ due to the accretion of BA parameterization for different values of $w_m$. We have taken three sets of $w_0$ and $w_1$ values to draw this graph. In figure 1a, $w_0= -0.98$ and $w_1 = 0.35$ in figure 1b, $w_0= -0.98$ and $w_1 = 0.16$ in figure 1c, $(w_0, w_1)=(-0.98,-0.01)$. The values of $w_0$ and $w_1$ are taken from the article \cite{Promila_thesis_2021}.}
\end{center}
\end{figure}

$\lambda_1^{BA}$ is determined by numerical integration and another numerical integration from a large distance to $x_c$ helped us to plot $log_{10}\left(\frac{M}{M_0}\right)$ in fig 3a-c. BAZS accretion curves are found different than CPL and JBP cases. Clearly, two phases of accretion are found. Upto a transition value of redshift, $z_t$, mass increases slowly. As we enter in the area $z<z_t$, a rapid increase is followed. This rapidness is again larger once the matter contribution possesses matter EoS $w_m<0$.
\subsection{Accretion of Kehagias-Sfetsos Black hole with Dark Energy Following Barboza-Alcaniz-Zhu-Silva Redshift Parameterization (BAZS) Parameterization}
Solving the equation of continuity \eqref{conservation eqn for fluid} for the BAZS EoS \eqref{EoS of BAZS}, we find the energy density of BAZS type DE as, 
\begin{equation}\label{rho phi0 of BAZS}
\rho^{BAZS}_{\phi}(z) = \rho^{BAZS}_{\phi 0}(1+z)^{\frac{3\{w_{\beta}+\beta(1+w^{BAZS}_0)\}}{\beta}}exp\left[\frac{3w_{\beta}}{\beta^2}\{(1+z)^{-\beta} -1\}\right]~~~~,
\end{equation}
where $\rho_{\phi 0}^{BAZS}$ is the DE density at present epoch. Hence the total energy density of universe filled up by BAZS type candidate is calculated as,
\begin{equation}\label{rho tot of BAZS}
\rho^{BAZS}_{tot}(z) = 3H_0^2 \left[\Omega_m (1+z)^3 + \Omega^{BAZS}_{\phi 0}(1+z)^{\frac{3\{w_{\beta}+\beta(1+w^{BAZS}_0)\}}{\beta}}exp\left[\frac{3w_{\beta}}{\beta^2}\{(1+z)^{-\beta} -1\}\right]\right]~~~~.
\end{equation}
So, putting the equation \eqref{rho tot of BAZS} in the equation \eqref{KS BH in HL gravity} we get for BAZS parameterization,
$$M^{BAZS}=~~~~~~~~~~~~~~~~~~~~~~~~~~~~~~~~~~~~~~~~~~~~~~~~~~$$
\begin{equation}\frac{M_0^{BAZS}}{1+\frac{4 \pi \lambda^{BAZS}_1 M_0}{3} \mathop{\mathlarger{\mathlarger{\int}}}_{\rho_{\phi_0}^{BAZS}}^{\rho_{\phi}^{BAZS}}\frac{d\left(3H_0^2 \left[\Omega_m (1+z)^3 + \Omega^{BAZS}_{\phi 0}(1+z)^{\frac{3\{w_{\beta}+\beta(1+w^{BAZS}_0)\}}{\beta}}exp\left[\frac{3w_{\beta}}{\beta^2}\{(1+z)^{-\beta} -1\}\right]\right] \right)}{\sqrt{
\frac{\kappa^2}{12}\left[\left(3H_0^2 \left[\Omega_m (1+z)^3 + \Omega^{BAZS}_{\phi 0}(1+z)^{\frac{3\{w_{\beta}+\beta(1+w^{BAZS}_0)\}}{\beta}}exp\left[\frac{3w_{\beta}}{\beta^2}\{(1+z)^{-\beta} -1\}\right]\right] \right)-6k\mu^4 (1+z)^2-\frac{3k\kappa^2 \mu^2 (1+z)^4}{16} \right]}}}~~~,
\end{equation}
$M^{BAZS}$ is the BH mass in BAZS at an arbitrary redshift ($M^{BAZS}$) and $M_0~(M(z=0))^{BAZS}$ is the present mass of the black hole in BAZS.
\begin{figure}[h!]
\begin{center}
$$~~~~~~~~~~~~~~~~~~~~~~~~Fig- 4a~~~~~~~~~~~~~~~~~~~~~~~~~~~~~~~~~~~~Fig-4b~~~~~~~~~~~~~~~~~~~~~~~~~~~~~~~~~~~~~~~~~~~~
~~Fig-4c~~~~~~~~~~~~~~~~~~~~~~~~~~~~~~$$\\
\includegraphics[height=1.5in, width=2.1in]{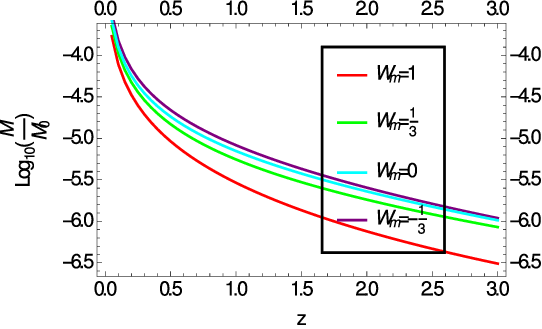}~~~
\includegraphics[height=1.5in, width=2.1in]{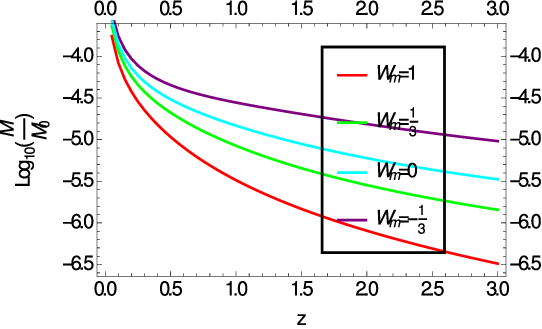}~~~~~
\includegraphics[height=1.5in, width=2.1in]{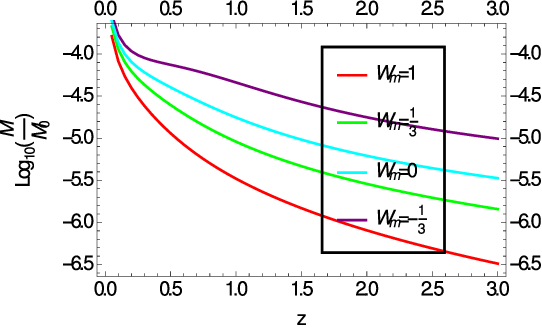}~~~~~\\
\caption{Variations in BH mass $M^{BAZS}$ against redshift $z$ due to the accretion of BAZS parameterization for different values of $w_m$. We have taken three sets of $w_0$ and $w_{\beta}$ values to draw this graph. In figure 1a, $w_0= -0.71$, $w_{\beta} = 1.56$ and $\beta = 0.5$ in figure 1b, $w_0= -0.68$, $w_{\beta} = -0.67$ and $\beta = -0.35$ in figure 1c, $(w_0, w_{\beta}, \beta )=(-0.54,~ {-3.58},~3.5)$.}
\end{center}
\end{figure}
We plot $log_{10}\left(\frac{M}{M_0}\right)$ for BAZS accretion as the previous cases. But the plots 4a-c are much different than previous cases. If $w_{\beta}$ is positive with a positive $\beta$, mass plots for different $w_m$ are followed not to differ much. They stayed only $10^{-6}$ to $10^{-6.5}$ orders of $M_0$ at $z\sim 3$. Once negative $w_{\beta}$ is chosen with small negative or positive $\beta$, huge differences for $w_m$ are noticed. For $w_0=-0.54$, $w_{\beta}=-3.58$, $\beta=3.5$ and $w_m=-\frac{1}{3}$, we follow a point of inflection to form around $z\sim 0.5$. This says $w_m=-\frac{1}{3}$ is an exotic curve which led the mass of the accreting black hole to reach a peak but failed to accumulate more mass at that stage. As time rolled, DE grew more dominating and we again follow hikes in $M$ as compared to $M_0$.
\newpage
\begin{figure}[h!]
\begin{center}
$$~~~~~~~~~~~~~~~~~~~~~~~~~~~~~~~~~~~~~Fig- 5a~~~~~~~~~~~~~~~~~~~~~~~~~~~~~~~~~~~~Fig-5b~~~~~~~~~~~~~~~~~~~~~~~~~~~~~~
~~~~~~~~~~~~~~$$\\
\includegraphics[height=2.5in, width=3.5in]{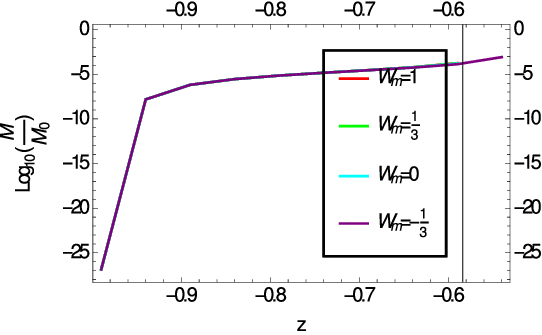}~~~~~
\includegraphics[height=2.5in, width=3.5in]{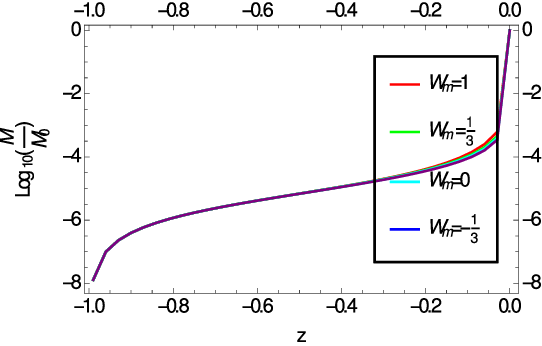}~~~~~\\
\caption{Future variations in BH mass $M^{CPL}$ against redshift $-1<z<0$ due to the accretion of CPL parameterization for different values of $w_m$ are plotted in figure 5a. Whereas fig 5b is the same profiles for BA}
\end{center}
\end{figure}
In figure 5a and 5b, we extrapolate the mass growth for CPL and BA parameterization universe. For this, we observe the mass of the central gravitating object to fall in future. Around $z\sim -1$, mass fall abruptly towards almost an order $\sim 10^{-25}$ of its present mass. Universe filled up with BA type fluid also forces a black hole to loose its mass and around $z\sim -1$, $10^{-8}$th order of present time black hole mass is left. These two results support the fate of black hole around a future singularity named ``Big Rip'' as mentioned by different articles \cite{Babichev_2004}. 
\section{Brief Discussion and Concluding Remarks}
While modification of gravity is almost a fashion by changing the Lagrangian of Einstein Hilbert action, Petr Horava proposed a quantum gravity where the `time' is treated to be more fundamental quantum quantity than the general relativity. If a detailed balanced Horava gravity is chosen with projectibility condition, a black hole solution is given in this gravity theory by Kehagias and Sfetsos. We have studied a Michel-like accretion around such a black hole. We have considered the universe to be filled of dark energy which homogeneously fill the cosmos all over and exerts negative pressure to generate the late time cosmic acceleration as speculated by different type Ia supernovae observations since 1998. Among several kinds of dark energy candidates, redshift parametrization models to support the $\Lambda CDM$ model at zero redshift and mimic other observations at different redshift models. In this article, four models - Chevallier-Polarski-Linder, Jassal-Bagala-Padmanabhan, Barboza-Alcaniz and Barboza-Alcaniz-Zhu-Silva are picked for accretion studies. 

When a fluid is accreted towards a black hole, transonic accretion may get formed if the fluid speed equals the sonic speed for that concerned accreting fluid at some finite radial distance. Dark energy candidate, being re-pulsars, nullifies attractive power of black hole and hence critical points are shifted to nearer points of the gravitating core. Now different candidate possesses different kind of repulsion. That too is different for different sets of free parameters. Hence not only for separate dark energy candidate but also for a particular candidate, every individual set of free parameters generates a new model and hence a distinct critical distance. This gives us different values of an important constant $\lambda_2$ which determines the growth of mass of the central gravitating object. Hence the mass accretion rate equation for each separate dark energy model is established and solved numerically to find the growing mass. To understand the situation deeply, we plot logarithm of the ratio of mass to present time mass for these models.

First, in Chevallier-Polarski-Linder universe we follow a Horava black hole to achieve $\sim 10^{-6.5}$ to $10^{-5}$ th part of its present day mass at around $z\sim 3$. Mass growth has two distinct rates, first slowly and then faster. For Jassal-Bagala-Padmanabhan filled universe, the mass growth is slower than Chevallier-Polarski-Linder. At present time, in both Chevallier-Polarski-Linder and Jassal-Bagala-Padmanabhan cases mass growth does not depend much on the dark matter's equation of state. In past ($z\sim 3$) $w_m=1$ and $w_m=-\frac{1}{3}$ cases differ large. Among them difference for Chevallier-Polarski-Linder is even larger than Jassal-Bagala-Padmanabhan. Black holes in Barboza-Alcaniz type dark energy filled universe is followed to grow steeply in the neighbourhood of $z\sim 0$. For Barboza-Alcaniz-Zhu-Silva with negative $w_0$ and $w_{\beta}$ with positive $\beta=3.5$, for a redshift range (in recent past) mass growth stops if $w_m=-\frac{1}{3}$.

Next question arose in our mind regarding the fate of such black holes in future. We follow, if our universe is supposed to be filled up by dark fluids of type Chevallier-Polarski-Linder or Barboza-Alcaniz, mass of a black hole will reduce in future due to accretion of surrounding dark fluids. At $z\sim -1$ mass is speculated to fall around $-25^{th}$ order of the present day mass of it. This indicates that in future, black holes sitting in Horava Lifshitz Gravity and a dark energy background evaporates to almost an infinitesimal mass especially for Chevallier-Polarski-Linder type of dark energy model. Barboza-Alcaniz model supports evaporation but not upto minute remnant. For Chevallier-Polarski-Linder case, before we may reach the ``Big Rip'', black holes are almost evaporated out such that the conflict regarding the fate of compact objects at future singularity even will not arise.  


\vspace{.7 in}

{\bf Acknowledgment:}
RB and FR thank IUCAA, Pune, India for providing Visiting Associateship. FR is also thankful to SERB, DST, DST, Govt. of India for financial support. 

\vspace{.1 in} 

{\bf Data Availability Statement : }
No data was generated or analyzed in this study.

\vspace{.1 in}

{\bf Conflict of Interest : }
There are no conflicts of interest.

\vspace{.1 in}

{\bf Funding Statement :}
There is no funding to report for this article.
\bibliographystyle{ieeetr} 

\bibliography{references}

\end{document}